\documentclass[journal,twocolumn,letterpaper]{IEEEJERM}
\usepackage{cite}

\ifCLASSINFOpdf
\else
\fi

\usepackage{times,amsmath,epsfig}
\usepackage{fancyhdr}
\usepackage{amsmath}
\usepackage{amsfonts}
\usepackage[latin1]{inputenc}
\usepackage{array}
\usepackage{graphicx}
\usepackage{url}
\usepackage{bm}
\usepackage{breqn}
\usepackage{xcolor}
\usepackage{soul}
\usepackage{amssymb}
\usepackage{color}
\usepackage{placeins}
\usepackage{float}
\usepackage{hyperref}
\usepackage{tabularx,colortbl}
\usepackage{bm}
\usepackage{comment}
\usepackage{tikz}
\usepackage{pgfplots}
\usepackage{pgfplotstable} 
\usepackage{siunitx} 
\usepackage[caption=false, font=footnotesize]{subfig}
\usepackage{cleveref}
\usepackage{adjustbox}

\pgfplotsset{compat=newest,
             every axis/.append style={
             label style={font=\footnotesize},
             tick label style={font=\footnotesize}}}
\usepgfplotslibrary{units} 

\DeclareMathAlphabet{\mathbfsf}{\encodingdefault}{\sfdefault}{bx}{n}
\DeclareMathAlphabet{\mathsfbfit}{OT1}{lmss}{bx}{sl}

\newcommand*{\sigmat}{\bar{\bar{\sigma}}}
\newcommand*{\br}{\bm{r}}
\newcommand*{\xik}{\xi_{\Gamma_k}}
\newcommand*{\xii}{\xi_{\Gamma_i}}
\newcommand*{\chit}{\bar{\bar\chi}}
\newcommand*{\Jk}{\bm{J}_{eq_k}}
\newcommand*{\Ji}{\bm{J}_{eq_i}}
\newcommand*{\nhat}{\hat{\bm{n}}}

\newcommand{\matI}{\mathbfsf{I}}
\newcommand{\mat}[1]{\mathsfbfit{#1}}
\newcommand{\dd}{\mathrm{d}}
\newcommand\copyrighttext{%
  \footnotesize \textcopyright 2020 IEEE. Personal use of this material is permitted.  Permission from IEEE must be obtained for all other uses, in any current or future media, including reprinting/republishing this material for advertising or promotional purposes, creating new collective works, for resale or redistribution to servers or lists, or reuse of any copyrighted component of this work in other works.}
\newcommand\copyrightnotice{%
\begin{tikzpicture}[remember picture,overlay]
\node[anchor=south,yshift=3pt] at (current page.south) {\fbox{\parbox{\dimexpr\textwidth-\fboxsep-\fboxrule\relax}{\copyrighttext}}};
\end{tikzpicture}%
}

\begin{document}

%
\title{A Hybrid Volume-Surface-Wire Integral Equation for the Anisotropic Forward Problem in Electroencephalography}

%
\author{Maxime~Y.~Monin,~\IEEEmembership{Student~Member,~IEEE,}
        Lyes~Rahmouni,
        Adrien~Merlini,~\IEEEmembership{Member,~IEEE,}
        and~Francesco~P.~Andriulli,~\IEEEmembership{Senior~Member,~IEEE}
}

\markboth{IEEE Journal of Electromagnetics, RF and Microwaves in Medicine and Biology}
{M.Y. Monin \MakeLowercase{\textit{et al.}}: A Hybrid Volume-Surface-Wire Integral Equation for the Anisotropic Forward Problem in Electroencephalography}

\twocolumn[
\begin{@twocolumnfalse}
  
\maketitle

\begin{abstract}
Solving the electroencephalography (EEG) forward problem is a fundamental step in a wide range of applications including biomedical imaging techniques based on inverse
source localization. State-of-the-art electromagnetic solvers resort to a computationally expensive volumetric discretization of the full head to account for its complex and heterogeneous electric profile. The more efficient, popular in biomedical imaging circles, but unfortunately oversimplifying Boundary Element Method (BEM) relies instead on a  piecewise-uniform approximation that severely curbs its application in high resolution EEGs. This contribution lifts the standard BEM contraints by treating the local anisotropies with adequate wire and thin volume integral equations that are tailored to specific structures of the fibrous white matter and the inhomogeneous skull. The proposed hybrid integral equation formulation thereby avoids the full volumetric discretization of the head medium and allows for a realistic and efficient BEM-like solution of the anisotropic EEG forward
problem. The accuracy and
flexibility of the proposed formulation is demonstrated through numerical
experiments involving both canonical and realistic MRI-based head models.
\end{abstract}


\end{@twocolumnfalse}]

{
  \renewcommand{\thefootnote}{}%
  \footnotetext[2]{This work was supported by the European Research Council (ERC) under the European Union's Horizon 2020 research and innovation programme (grant agreement No 724846, project 321). Data were provided in part by the Human Connectome Project, WU-Minn Consortium (Principal Investigators: David Van Essen and Kamil Ugurbil; 1U54MH091657) funded by the 16 NIH Institutes and Centers that support the NIH Blueprint for Neuroscience Research; and by the McDonnell Center for Systems Neuroscience at Washington University.}
  \footnotetext[3]{The authors are with the Department
of Electronics and Telecommunications, Politecnico di Torino, Turin, 10129, Italy (e-mail: maxime.monin@polito.it, lyes.rahmouni@polito.it, adrien.merlini@polito.it, francesco.andriulli@polito.it).}
}

\copyrightnotice
 
%
\IEEEpeerreviewmaketitle

\section{Introduction}
%
%
%
%
\IEEEPARstart{H}{igh}-resolution electroencephalography (EEG) is a staple neuroimaging tool
employed in a wide variety of biomedical applications ranging from clinical
diagnostics to research in neuroscience. An EEG system records the electric
potential on scalp electrodes generated by the brain neural activity at a high
temporal resolution and is both affordable and portable, which has facilitated
its broad adoption in the last few decades \cite{pfurtscheller2006mu,smith2005eeg,wendel2009eeg}. Compared to other imaging modalities, however, EEG imaging is sensitive to the smearing effects of
various head tissues, resulting in low spatial resolution. This challenge can
be tackled through advanced computational techniques
\cite{pascual1999review}, thus enabling precise analysis of deep neural
activity with EEG \cite{kiebel2006dynamic}.

The EEG forward problem is a characterization of the relationship between the
neural activity in the brain and the electric potential measured on the scalp.
Its solution is necessary for the associated inverse
problem \cite{grech2008review} that consists in inferring knowledge about brain
activity from EEG data. This characterization is challenging due to the complex nature of the human head. In particular, the skull
is made of varying thicknesses of soft bone and highly resistive hard bone layers \cite{sadleir2007modeling, papageorgakis2017patient}, whereas diffusion Magnetic Resonance Imaging (dMRI) shows that the brain white matter has a fiber-like structure of coherently oriented axon bundles connecting different brain regions \cite{le2012diffusion}. Overall, these fine structures make the head volume strongly inhomogeneous and anisotropic, and must be suitably taken into account as they have a strong influence on brain imaging accuracy \cite{lee2012regional, shahid2013numerical, vorwerk2014guideline}. In realistic applications, the forward problem is
solved using numerical techniques \cite{hallez2007review} derived
from full-wave or quasi-static differential and integral formulations of Maxwell's equations
\cite{bruno2006fdm,montes2014influence,lew2009accuracy,hamalainen1989realistic,kybic2005common,gomez2017icvsie}. State-of-the-art differential methods such as the Finite Element Method (FEM) can easily incorporate the local variations of tissue conductivity, but rely on a computationally expensive volumetric discretization of the entire head. The Boundary Element Method (BEM) \cite{niedermeyer2005electroencephalography, brette2012handbook} is a popular alternative which reformulates the forward problem with surface integrals on the boundaries of the head compartments, meaning that the linear systems to solve are considerably smaller. The BEM is numerically stable and can be further augmented using acceleration techniques such as the Fast Multipole Method (FMM) to reach great computational efficiency \cite{liu2009fast,makarov2018quasi,htet2019comparative}. Unfortunately, the standard BEM requires the head medium to be piecewise-uniform, and is consequently unable to model the anisotropies and inhomogeneities of the skull and of the white matter. Therefore, this intrinsic assumption drastically reduces its applicability.

A BEM formulation that
accounts for the skull anisotropy has recently been proposed
\cite{rahmouni2017two} but relies on a volumetric discretization of the whole
head. On the other hand, the white matter fibers can be quantitatively reconstructed from tractography algorithms applied on dMRI data \cite{Yamada2009MRTA,wandell2016clarifying}. This
representation is promising in that it describes
white matter in a much more effective and precise way than when relying on a
complete volume discretization which is independent of the fiber structure.
The thin fiber geometry is a common structure in high frequency electromagnetic problems \cite{wilton2006evaluation} and has been studied in a few contributions for bioelectromagnetic problems \cite{olivi2011handling, pillain2016line, makarov2016modeling, rahmouni2019boundary}. Tractography has also
been used in transcranial brain stimulation studies \cite{nummenmaa2014targeting,
opitz2011brain}, although the fiber structure was neglected in the
forward model construction. At present and to the best of our
knowledge, no contribution has been proposed to solve the anisotropic forward problem without resorting to a full volumetric discretization of the head.

In this paper, we address this issue by modifying and complementing the standard BEM  (2D) equations with adequate anisotropy-handling wire (1D) equations for the white matter and volume (3D) equations for the skull. The resulting new hybrid integral formulation is effectively tailored to the EEG forward problem as every head tissue is suitably discretized according to its electrical properties.
The integration of tractography algorithms for white matter conductivity profiling is obtained with an electrically coherent derivation of the fiber parameters, enabling the computation of a multimodal MRI consistent solution of the anisotropic EEG forward problem. The validity of the proposed scheme
is confirmed by numerical experiments which demonstrate its practical relevance.
Very preliminary results from this work were presented at the conferences \cite{monin2018hybrid, moninfutureiceaa}.

The background and notation required throughout this paper are introduced in
\Cref{sec:background} and the new formulation is developed along with its
discretization in \Cref{sec:new_eq}. Diffusion MRI considerations for white matter modeling are discussed in \Cref{sec:wire_params}, followed by numerical
results in
\Cref{sec:numerical}. Conclusive remarks are presented in
\Cref{sec:conclusion}.

\section{Background and Notations} \label{sec:background}

In the following, the head medium $\Omega$ will be modeled by nested layers $\Omega_i$ ($\Omega =
\bigcup_{i=1}^N \Omega_i$) with respective external boundaries $\Gamma_i$
($\Gamma = \bigcup_{i=1}^N \Gamma_i)$ and scalar background conductivities
$\sigma_i$, typically accounting for brain, skull and scalp \cite{brette2012handbook}. We assume a nested
structure for the sake of simplicity and readability but the formulation can be extended to arbitrary
geometries. In all generality, the  conductivity is expressed as a symmetric positive semidefinite spatial tensor
\begin{equation}
\label{eq:anisocond}
    \sigmat(\bm{r}) = U(\bm{r})
    \begin{bmatrix}
    \sigma_{v_1}(\br) & 0 & 0 \\
    0 & \sigma_{v_2}(\br) & 0 \\
    0 & 0 & \sigma_{v_3}(\br)
  \end{bmatrix}U^T(\bm{r}),
\end{equation}
where $\sigma_{v_k}$ are the conductivity eigenvalues and the unitary matrix $U$ contains the associated eigenvectors. Classical BEM assumes that in each compartment, the conductivity is reasonably approximated by the background conductivity so that $\sigmat(\br\in\Omega_i) \approx \sigma_i\matI$, where $\matI$
is the identity matrix. In practice, this is poorly verified in the skull, and in the fibrous white matter compartment $\Omega_{i_w}$, in which ions flow preferentially along the local fiber direction \textbf{$\hat{\bm{l}}(\br)$} so that $\sigmat = \sigma_{i_w} \matI + (\sigma_{l}-\sigma_{i_w}) \hat{\bm{l}} \hat{\bm{l}}^T\,,$ where the longitudinal conductivity $\sigma_l$ is approximately ten times higher than the background conductivity $\sigma_{i_w}$ \cite{NICHOLSON1965386, lozano2013brain}. The external region $\Omega_{N+1}$ represents
air and it is not conductive, therefore $\sigma_{N+1}=0$. In this
framework, solving the forward problem amounts to determining the unknown
electric potential $\phi(\br)$ on the scalp surface $\Gamma_N$ generated by a
known source located at $\br_0\in\Omega$ inducing a primary current
$\bm{J}_p(\br)$. In the quasi-static regime, these quantities are linked by
Poisson's equation
\begin{equation}
\nabla\cdot\left( \sigmat\nabla\phi \right)=\nabla\cdot \bm{J}_s,\quad \br\in\Omega\,,
\label{eq:poisson}
\end{equation}
with the boundary conditions
\begin{subequations}
\label{eq:bc}
\begin{align}
\phi|^-_i&=\phi|^+_i\,,\quad \br\in\Gamma_i\,, \label{eq:bc1}\\
\nhat\cdot\bar{\bar{\sigma}}\nabla\phi|^-_i &=
\nhat\cdot\bar{\bar{\sigma}}\nabla\phi|^+_i\,,\quad
\br\in\Gamma_{i<N}\,, \label{eq:bc2}\\
\nhat\cdot\bar{\bar{\sigma}}\nabla\phi&=0\,,\quad \br\in\Gamma_N\,, \label{eq:bc3}
\end{align}
\end{subequations}
where the dependency in $\br$ has been dropped for the sake of brevity, $f|^+_i$ and $f|^-_i$ denote respectively the exterior and interior
limits of a function $f$ on surface $\Gamma_i$,
$\nhat$ is the unitary outward normal vector of $\Gamma_i$ and $\bm{J}_s$ is the primary
source term; in the current setting $\bm{J}_s=\bm{J}_p$. In the standard single-layer BEM framework \cite{kybic2005common}, \eqref{eq:poisson} and \eqref{eq:bc} are reformulated in terms of integral equations and we define the surface and volume operators
\begin{subequations}
\begin{align}
    (\mathcal{S} f_s)(\br)&=
    \int_{S}G\left(\br,\br'\right)f_s\left(\br'\right)\dd S'\,, \quad \br\in\Omega\,,\\
    (\mathcal{D}^* f_s)(\br)&= \int_{S}\nabla_n
    G(\br,\br')f_s(\br)\dd S'\,, \quad \br\in\Gamma\,,\\
    (\mathcal{S}^*_v \bm{f}_v)(\bm{r})&= \int_{V} G(\br,\br')\nabla' \cdot
    \bm{f}_v(\bm{r}')\dd V'\,, \quad \br\in\Omega\,,\\
    (\mathcal{D}^*_v \bm{f}_v)(\bm{r})&= \int_{V}\nabla_n G(\bm{r},\bm{r}')\nabla' \cdot \bm{f}_v\left(\bm{r}'\right)\dd V'\,, \quad \br\in\Gamma\,,
\end{align}
\end{subequations}
where $\nabla_n=\hat{\bm{n}} \cdot \nabla$, $f_s$ (resp. $\bm{f}_v$) is a scalar (resp. vector) function defined on a
surface $S$ (resp. volume $V$), and $G(\bm{r},\bm{r}')=1/4\pi\|\bm{r}-\bm{r}'\|$ is the static Green's function. To describe the disparity between background isotropic and actual conductivity, we define in each compartment $\Omega_i$ the conductivity contrast $\chit_i(\br)=(\sigma_i\matI -\sigmat(\br))\sigmat^{-1}(\br)$.
The exterior domain is uniform, so we can write $\chit_{N+1}=0$ without loss of generality. Since the conductivity contrast is not negligible in realistic skull and white matter models, the applicability of a pure surface integral equation method is compromised.
\section{A New Hybrid Integral Equation Formulation} \label{sec:new_eq}
\subsection{Integral equations for surface contributions}\label{subs:hnuc}
In all generality, Poisson's equation \eqref{eq:poisson} in
each compartment can be rewritten
\begin{equation}\label{eq:poissonnonuni}
\sigma_{i}\Delta\phi(\br)=\nabla\cdot (\bm{J}_p(\br)+\chit_i(\br)\sigmat(\br)\nabla\phi(\br)),\quad \br\in\Omega_{i}.
\end{equation}
Hence, a non-uniform conductivity in $\Omega_i$ is equivalent to a volumetric source
generating a current $\Ji=\chit_i\sigmat\nabla\phi$ in a medium with arbitrary
uniform background conductivity $\sigma_i$. The left-hand side of \eqref{eq:poissonnonuni} implies that the anisotropic forward problem can be recasted as a piecewise-homogeneous one (with conductivity $\sigma_i$ in each compartment) provided that the source term $\bm{J}_s = \bm{J}_p+\sum_k \Jk$ now contains the equivalent current contributions. Therefore, following the classical BEM \cite{kybic2005common}, we define the single-layer potential 
\begin{equation} \label{eq:slpot}
    \phi_{SL} = \sum_{k=1}^N\mathcal{S}\xik = \phi + \tfrac{1}{\sigma_p}\mathcal{S}^*_v\bm{J}_p + \sum_{k=1}^N\tfrac{1}{\sigma_k}\mathcal{S}^*_v\Jk\,,
\end{equation}
where $\xii = [\nabla_n \phi]_{\Gamma_i}$. On each interface $\Gamma_i$, the boundary condition \eqref{eq:bc2} imposes
\begin{alignat}{2}\label{eq:devsurf}
    [\nhat\cdot\sigmat\nabla\phi]_{\Gamma_i}=[\sigma_i\nabla_n\phi]_{\Gamma_i} - [\nhat\cdot\Ji]_{\Gamma_i}=0.
\end{alignat}
Using \eqref{eq:slpot}, the term $[\sigma_i\nabla_n\phi]_{\Gamma_i}$ can be broken into
\begin{align}\label{eq:devsurfbroken}
    [\sigma_i\nabla_n \phi_{SL}]_{\Gamma_i} &= \tfrac{\sigma_i+\sigma_{i+1}}{2}\xi_{\Gamma_i} + (\sigma_i-\sigma_{i+1})\sum_k\mathcal{D}^*\xi_{\Gamma_k}\,, \nonumber \\
    [\sigma_i\nabla_n\tfrac{1}{\sigma_p}\mathcal{S}^*_v\bm{J}_p]_{\Gamma_i} &= \tfrac{\sigma_{i} - \sigma_{i+1}}{\sigma_p}\mathcal{D}^*_v\bm{J}_p\,, \nonumber \\
    \sum_k[\tfrac{\sigma_i}{\sigma_k}\nabla_n\mathcal{S}^*_v\Jk]_{\Gamma_i} &= (\sigma_{i} - \sigma_{i+1})\sum_k\tfrac{1}{\sigma_k}\mathcal{{D}}^*_v\Jk \nonumber \\ + \nhat\cdot &(\tfrac{\sigma_i+\sigma_{i+1}}{2\sigma_i}\Ji - \tfrac{\sigma_i+\sigma_{i+1}}{2\sigma_{i+1}}\bm{J}_{eq_{i+1}})\,,
\end{align}
where we used the fact that the conductivity contrast, and therefore the equivalent currents $\Ji$ and $\bm{J}_{eq_{i+1}}$ are discontinuous across $\Gamma_i$, so that the double-layer volume operator
$\mathcal{D}^*_v$ in \eqref{eq:devsurfbroken} gives rise to the discontinuity 
\begin{align}\label{eq:jumpm}
    [\mathcal{D}^*_v\Ji]_{\Gamma_i}&=\nhat\cdot\Ji\,, \quad \br\in\Gamma_i\,, \\
    [\mathcal{D}^*_v\bm{J}_{eq_{i+1}}]_{\Gamma_i}&=-\nhat\cdot\bm{J}_{eq_{i+1}}\,, \quad \br\in\Gamma_i\,.
\end{align}
Substituting \eqref{eq:devsurfbroken} in \eqref{eq:devsurf}, the equation on each surface $\Gamma_i$ is
\begin{multline}\label{eq:surf}
     \tfrac{\sigma_i+\sigma_{i+1}}{2(\sigma_{i+1}-\sigma_i)}\xi_{\Gamma_i} + \nhat\cdot(\tfrac{1}{2\sigma_{i+1}}\bm{J}_{eq_{i+1}}
    - \tfrac{1}{2\sigma_i}\Ji)\\
    - \sum_k [\mathcal{D}^*\xi_{\Gamma_k} - \tfrac{1}{\sigma_k}\mathcal{{D}}^*_v \bm{J}_{eq_k}] =-\tfrac{1}{\sigma_p}\mathcal{D}^*_v\bm{J}_p\,.
\end{multline}
To obtain a discrete system, the surface unknowns $\xii$ are expanded with pyramidal basis functions $\xi_{\Gamma}(\br) = \sum_j \alpha_{p_j} p_{j}(\br)$,
where $p_{j}$ is non-zero in any triangle $T_{jkl}$ defined by vertices $\br_j,
\br_k, \br_l$
\begin{equation}
    p_{j}(\br)=
    \begin{cases}
        \frac{|(\br-\br_l) \times (\br_k-\br_l)|}{|(\br_j-\br_l) \times (\br_k-\br_l)|}, & \br\in T_{jkl}\,, \\
        0, & \text{otherwise}.
    \end{cases}
\end{equation}
Following a Galerkin approach the surface equations are tested with the same basis functions, resulting in the adjoint double-layer and sparse Gram self-terms 
\begin{subequations}
\begin{align}
    (\mat{G}_{pp})_{mn}&= \tfrac{\sigma_{i_n}+\sigma_{i_n+1}}{2(\sigma_{i_n+1}-\sigma_{i_n})}\langle p_m, p_n \rangle_\Gamma, \\
    (\mat{D}^*_{pp})_{mn}&= \langle p_m, \mathcal{D}^*p_n \rangle_\Gamma\,,
\end{align}
\end{subequations}
in which $\langle a,b \rangle _\kappa=\int_{\kappa}
a(\br) b(\br)\dd \kappa$, and $m$ and $n$ denote a testing row and a basis column, respectively.

Contrary to $\bm{J}_p$, the equivalent currents are secondary sources in that they arise from the primary activity and are dependent on the media. Therefore they must be determined by coupling \eqref{eq:surf} with additional equations.

\subsection{Integral equations for volume contributions}\label{subs:vie}
By applying the gradient operator to \eqref{eq:slpot} in the inhomogeneous regions, we obtain a new set of independent equations
\begin{multline}\label{eq:volume}
   -(\sigma_i\matI-\sigmat)^{-1}\bm{J}_{eq_i} +  \sum_k\left[\nabla\mathcal{S}\xi_{\Gamma_k}-\frac{1}{\sigma_k}\nabla\mathcal{S}^*_v\bm{J}_{eq_k}\right] \\ = \frac{1}{\sigma_p}\nabla\mathcal{S}^*_v\bm{J}_p\,.
\end{multline}
Note that \eqref{eq:volume} is valid when the conductivity contrast is not null, so that the volume equations requires the discretization of the inhomogeneous domain only. The equivalent volume currents are expanded with Schaubert-Wilton-Glisson (SWG) basis functions \cite{schaubert1984tetrahedral} $\bm{J}_{eq}(\br) = \chit(\br)\sum_j \alpha_{s_j} \bm{s}_{j}(\br)$,
where $\bm{s}_j$ is defined on the tetrahedron pair $T_j^\pm$ with volumes $V_j^\pm$, opposite vertices $\br_j^\pm$ and common triangle $t_j$ and area $a_j$, as
\begin{equation}
    \bm{s}_j(\br)= 
\begin{cases}
    \frac{a_j}{3V_j^+}(\br-\br_j^+)\,,& \br\in T_j^+,\\
    -\frac{a_j}{3V_j^-}(\br-\br_j^-)\,,& \br\in T_j^-,\\
    \bm{0}\,,& \text{otherwise}.
\end{cases}
\end{equation}
This choice of basis function enforces the continuity of the current density across each pair of tetrahedra, and also allows integration by parts to transfer the source gradient to the testing side when computing the entries of the volume matrix. The volume self terms are 
\begin{subequations}
\begin{align}
    (\mat{G_v}_{ss})_{mn}&= \langle \bm{s}_m, \chit^{-1}\bm{s}_n \rangle_\Omega, \\
    (\mat{S^*_v}_{ss})_{mn}&= \tfrac{1}{\sigma_{i_n}}\langle \bm{s}_m,
    \nabla\mathcal{S}_v^*\chit\bm{s}_n \rangle_\Omega \nonumber \\
    &= -\tfrac{1}{\sigma_{i_n}}\langle \nabla \cdot \bm{s}_m,
    \mathcal{S}_v^*\chit\bm{s}_n \rangle_\Omega\,, \label{eq:intbyparts1}
\end{align}
\end{subequations}
where, for implementation purposes, the differential operator on the source integral is transferred to the testing side via integration by parts and Gauss divergence theorem. In particular, on the boundary of the tetrahedral mesh, the divergence of the half SWG (in both test and source integral) gives rise to a surface integral on its defining triangle $t_m$, e.g.\
\begin{align} \label{eq:intbyparts2}
    ({\mat{S^*_v}}_{ss})_{mn} &= \tfrac{1}{\sigma_{i_n}}\langle \nhat \cdot \bm{s}_m,
    \mathcal{S}_v^*\chit\bm{s}_n \rangle_{\partial\Omega}\ -\tfrac{1}{\sigma_{i_n}}\langle \nabla \cdot \bm{s}_m,
    \mathcal{S}_v^*\chit\bm{s}_n \rangle_\Omega\ \nonumber
    \\ &= \tfrac{1}{\sigma_{i_n}}\int_{t_m}\mathcal{S}_v^*\chit\bm{s}_n(\br)\dd S -\tfrac{a_m}{\sigma_{i_n}V_m^+}\int_{T_m^+}\mathcal{S}_v^*\chit\bm{s}_n(\br)\dd V
\end{align}

for a half testing SWG function $\bm{s}_m$.
\subsection{Integral equations for fiber contributions}\label{subs:wie}
The white matter compartment $\Omega_{i_w}$ is locally more conductive along a specific direction $\hat{\bm{l}}(\br)$, so that the non-uniform region is locally modeled as a thin cylindrical fiber. The fiber contrast $\chit_{i_w}$ is a projection on $\hat{\bm{l}}$, which implies that the volume
unknown along a fiber, $\bm{J}_{eq_{i_w}} = (\sigma_{i_w} -
\sigma_l)\frac{\partial \phi}{\partial l}\hat{\bm{l}} = J_{eq_{i_w}}(l)\hat{\bm{l}}\,,$ is a scalar function along the fiber direction. Hence, by taking the derivative of \eqref{eq:slpot} along $\hat{\bm{l}}$ and with further manipulations, we obtain the wire integral equations
\begin{multline}\label{eq:wire}
    -\frac{1}{\sigma_{i_w}-\sigma_l}J_{eq_{i_w}} + \sum_k[\nabla_l\mathcal{S}\xi_{\Gamma_k}-\tfrac{1}{\sigma_k}\nabla_l\mathcal{S}^*_v\bm{J}_{eq_k}] \\ = \tfrac{1}{\sigma_p}\nabla_l\mathcal{S}^*_v\bm{J}_p,\quad \chit_{i_w}(\br) \neq 0\,.
\end{multline}
Following the fiber conductivity model, the fiber equivalent currents in the white matter are expanded using one-dimensional basis functions with longitudinal
orientation and constant value on the fiber cross-section. Piecewise linear (hat) basis functions
defined on consecutive fiber segments $s^-=[\bm{r}_{j-1}; \bm{r}_{j}]$ and
$s^+=[\bm{r}_{j}; \bm{r}_{j+1}]$ as 
\begin{equation}
    \bm{h}_j(\br)= 
\begin{cases}
    \frac{1}{|\br_j - \br_{j-1}|}(\br-\bm{r}_{j-1})\,,& \br\in s^-,\\
    \frac{1}{|\br_{j+1} - \br_{j}|}(\br_{j+1}-\br)\,,& \br\in s^+,\\
    \bm{0}\,,& \text{otherwise},
\end{cases}
\end{equation}
automatically enforce continuity of the current density across segments and, similarly to the SWG functions in the 3D case, are divergence conforming, which allows the transfer of the gradient in the source integral to the testing function. The fiber self terms are
\begin{subequations}
\begin{align}
    (\mat{G_v}_{hh})_{mn}&= \langle \bm{h}_m, \chit^{-1}\bm{h}_n \rangle_\Omega, \\
    (\mat{S^*_v}_{hh})_{mn}&= \tfrac{1}{\sigma_{i_n}}\langle \bm{h}_m,
    \nabla\mathcal{S}_v^*\chit\bm{h}_n \rangle_\Omega\,.
\end{align}
\end{subequations}
The $\mat{S^*_v}_{hh}$ operator can be efficiently computed with one-dimensional integrals along the fiber direction or semi-analytically for far and near interactions, respectively.
\subsection{Coupling terms}
As a direct result of the coupling between the surface, volume and fiber unknowns in the associated equations \eqref{eq:surf}, \eqref{eq:volume} and \eqref{eq:wire}, the hybrid system matrix contains non-zero off-diagonal terms.
\begin{subequations}
\begin{align}
    (\mat{G}_{ps})_{mn}&= \tfrac{1}{2\sigma_{i_n}}\langle p_m,\nhat\cdot\chit\bm{s}_n \rangle_\Gamma, \\
    (\mat{D^*_v}_{ps})_{mn}&= \tfrac{1}{\sigma_{i_n}}\langle p_m, \mathcal{D}_v^*\chit\bm{s}_n \rangle_\Gamma, \\
    (\mat{D^*_v}_{ph})_{mn}&= \tfrac{1}{\sigma_{i_n}}\langle p_m, \mathcal{D}_v^*\chit\bm{h}_n \rangle_\Gamma, \\
    (\mat{S}_{sp})_{mn}&= \langle \bm{s}_m, \nabla\mathcal{S}p_n \rangle_\Omega, \\
    (\mat{S}_{hp})_{mn}&= \langle \bm{h}_m, \nabla\mathcal{S}p_n \rangle_\Omega, \\
    (\mat{S^*_v}_{hs})_{mn}&= \tfrac{1}{\sigma_{i_n}}\langle \bm{h}_m,
    \nabla\mathcal{S}_v^*\chit\bm{s}_n \rangle_\Omega, \\
    (\mat{S^*_v}_{sh})_{mn}&= \tfrac{1}{\sigma_{i_n}}\langle \bm{s}_m,
    \nabla\mathcal{S}_v^*\chit\bm{h}_n \rangle_\Omega.
\end{align}
\end{subequations}
The mixed Gram operators between the surface and volume unknowns are a consequence of the fact that a volume-discretized compartment $\Omega_i$ is bounded by the surface-discretized interfaces $\Gamma_i$ and $\Gamma_{i-1}$. Like the self terms in \eqref{eq:intbyparts1} and \eqref{eq:intbyparts2}, the source differential operators are transferred to the divergence-conforming testing side and computed using similar strategies as in \cite{graglia1993numerical,jarvenpaa2003singularity}.

\subsection{Solution of the resulting anisotropic forward problem}

The right-hand side (RHS) of the discretized equations is
\begin{subequations}
\begin{align}
    (\mat{v}_{p})_{m}&= -\tfrac{1}{\sigma_s}\langle p_m, \mathcal{D}^*_v\bm{J}_p \rangle_\Gamma, \\
    (\mat{v}_{h})_{m}&= -\tfrac{1}{\sigma_p}\langle \bm{h}_m, \nabla\mathcal{S}^*_v\bm{J}_p \rangle_\Omega = \tfrac{1}{\sigma_p}\langle \nabla\cdot\bm{h}_m, \mathcal{S}^*_v\bm{J}_p \rangle_\Omega, \\
    (\mat{v}_{s})_{m}&= -\tfrac{1}{\sigma_p}\langle \bm{s}_m, \nabla\mathcal{S}^*_v\bm{J}_p \rangle_\Omega = \tfrac{1}{\sigma_p}\langle \nabla\cdot\bm{s}_m, \mathcal{S}^*_v\bm{J}_p \rangle_\Omega.
\end{align}
\end{subequations}
The hybrid matrix system
\begin{gather}\label{eq:fullmatrix}
 \small{\begin{bmatrix} -\mat{G}_{pp}+\mat{D}^*_{pp} & -\mat{D^*_v}_{ph} & \mat{G}_{ps}-\mat{D^*_v}_{ps} \\ -\mat{S}_{hp} & \mat{G_v}_{hh}+\mat{S^*_v}_{hh} &  \mat{S^*_v}_{hs} \\ -\mat{S}_{sp} & \mat{S^*_v}_{sh} & \mat{G_v}_{ss}+\mat{S^*_v}_{ss}\end{bmatrix}
 \begin{bmatrix}
 \alpha_p\\ \alpha_h \\ \alpha_s
 \end{bmatrix}
 =
  \begin{bmatrix}
   \mat{v}_{p} \\
   \mat{v}_{h} \\
   \mat{v}_{s}
   \end{bmatrix}}
\end{gather}
has a well-known singularity in the surface-surface (top-left) block which is removed by deflation \cite{chan1984deflated}. After inverting the final system, the potential in $\Omega$ is finally computed as
\begin{multline}\label{eq:potfinal}
    \phi(\br) = \sum_i \alpha_{p_i}\mathcal{S}p_i(\br)-\sum_j \alpha_{h_j}\mathcal{S}^*_v\chit\bm{h}_j(\br) \\ - \sum_k \alpha_{s_k}\mathcal{S}^*_v\chit\bm{s}_k(\br) -\tfrac{1}{\sigma_p}\mathcal{S}^*_v\bm{J}_p(\br).
\end{multline}
\section{A New Diffusion MRI Consistent White Matter Model}\label{sec:wire_params}
Although tractography provides the mesh for the fibers, their radius, longitudinal and background conductivity must be determined. To this end, we propose next a new procedure to enable and validate the proposed tractography-derived conductivity model of white matter. First the fiber radius is adjusted to match a known volume of white matter (approximately \SI{450}{\centi\meter\cubed} \cite{zhang2000universal}). 
Then, for a given fiber conductivity, an effective homogenized conductivity of the brain $\bar{\bar{\sigma}}_e$ is determined by solving a homogenization problem with integral equations. Denoting $\Omega_{1}$ the brain compartment (which contains the fibers),
$\bar{\bar{\sigma}}_e$ is defined as $\langle \bm{J} \rangle = \bar{\bar{\sigma}}_e \langle \bm{E} \rangle$, where $\bm{J}$ is the current density, $\bm{E}=-\nabla\phi$ is the electric
field and $\langle \cdot \rangle = V^{-1}_{\Omega_{1}}\int_{\Omega_{1}}{\cdot \
d\Omega_{1}}$ denotes the volume average over the brain. By imposing a normal current density $\bm{J}_0$ on the boundary, that is
\begin{equation}\label{eq:bcJ0}
    \nhat\cdot\sigmat\nabla\phi = \nhat\cdot\bm{J}_0,\ \br\in\Gamma_{1}, 
\end{equation}
and using Gauss divergence theorem, the volume averaged current density
$\langle \bm{J} \rangle$ is equal to $\bm{J}_0$. Furthermore, a forward problem
for the Poisson's equation in the absence of primary source in the volume, $\nabla\cdot\sigmat\nabla\phi= 0,\ \br\in\Omega_{1},$ and complemented with the boundary condition \eqref{eq:bcJ0} can be solved with
our hybrid formulation to obtain the scalar potential $\phi$ on the boundary
$\Gamma_1$, as given by \eqref{eq:potfinal}. Using again Gauss divergence
theorem, the computation of $\langle \bm{E} \rangle$ is given by
\begin{equation}
    \langle \bm{E}^{(\bm{J}_0)} \rangle = -\frac{1}{V_{\Omega_{1}}}\int_{\Gamma_{1}}{\phi(\br) \nhat(\br) d\Gamma_{1}},
\end{equation}
where the $\bm{J}_0$ superscript indicates the dependency on the imposed current.
Thus, by solving this forward problem three times for $\bm{J}_0 \in \{\bm{x},
\bm{y}, \bm{z}\}$ (the unit cartesian vectors), the equivalent conductivity is
obtained as
\begin{equation}
    \bar{\bar{\sigma}}_e =
    \begin{bmatrix}
    \langle \bm{E}^{(\bm{x})} \rangle_x & 0 & 0 \\
    0 & \langle \bm{E}^{(\bm{y})} \rangle_y & 0 \\
    0 & 0 & \langle \bm{E}^{(\bm{z})} \rangle_z
  \end{bmatrix}^{-1},
\end{equation}
where the subscripts on vectors $\langle \bm{E}^{(\bm{J}_0)} \rangle$ denote
their cartesian components. As $\bar{\bar{\sigma}}_e$ satisfies
Ohm's law averaged over the volume, it corresponds to the electrically
equivalent average (homogenized) conductivity of the brain for the specified boundary conditions \eqref{eq:bcJ0}. This value can then be compared with brain conductivity values reported in literature to ensure that the obtained tractography-based conductivity model of the white matter is realistic. This will be demonstrated in the numerical results.

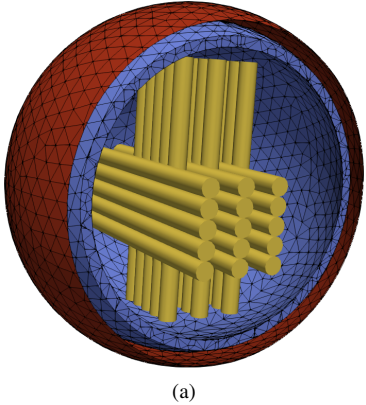
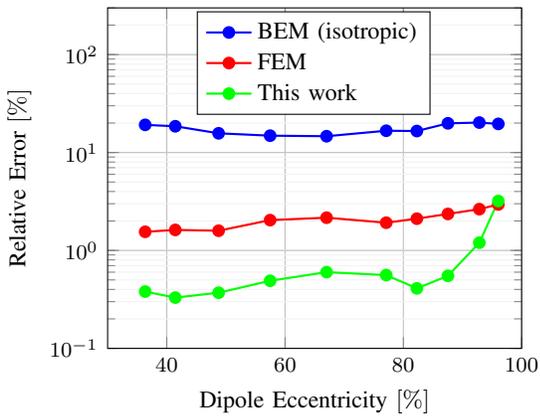
\begin{figure}
    \centering
    \trimbox{0pt 0pt 0pt 30pt}{\subfloat[]{\includegraphics[width=0.55\columnwidth]{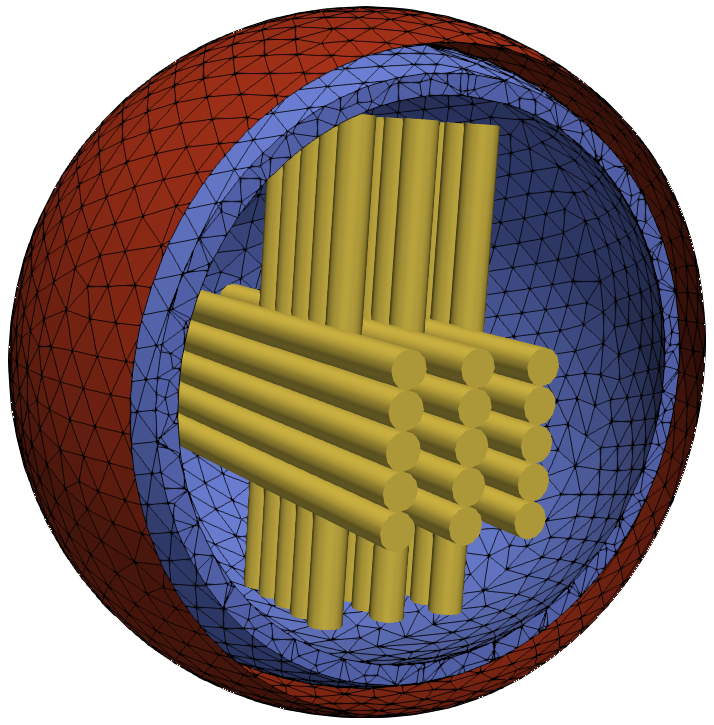}\label{fig:sph_svwA}}}
    
    \subfloat[]{%
    \begin{tikzpicture}
      \begin{axis}[%
          width=0.8\columnwidth, 
          grid=both,
          grid style={, draw=gray!10},
          major grid style={,draw=gray!50},
          label style={font=\small},
          xlabel=Dipole Eccentricity, 
          ylabel=Relative Error,
          x unit=\si{\%}, 
          y unit=\si{\%},
          ymode = log,
          ytick={1e-1, 1e0, 1e1, 1e2, 1e3},
          ymin = 1e-1,
          ymax = 3e2,
          xmin = 30,
          xmax = 100,
          xtick={0, 20, 40, 60, 80, 100},
          legend style={at={(0.78,0.99)},anchor= north east,font=\small},
          legend cell align={left}
        ]
        \addplot [mark=*, blue, thick]
        table[x=exc,y=bem_err,col sep=comma] {data/err_svw_sph.txt}; 
        \addplot [mark=*, red, thick]
        table[x=exc,y=femft_err,col sep=comma] {data/err_svw_sph.txt}; 
        \addplot [mark=*, green, thick]
        table[x=exc,y=hybrid_err,col sep=comma] {data/err_svw_sph.txt};
        \legend{BEM (isotropic),FEM,This work}
      \end{axis}
    \end{tikzpicture}\label{fig:sph_svw_err}}
    \caption{(a) Spherical head geometry with anisotropic and inhomogeneous brain and skull compartments; (b) relative error of the proposed method, OpenMEEG BEM and FieldTrip FEM.}
    \label{fig:sph}
\end{figure}
\begin{table*}[ht]

\caption{Partial and total timing of the proposed method and FEM in a realistic scenario}
\begin{center}
\begin{tabular}{|l||c|c|c|c|c|c|}
\hline
      & Setup time & Time per RHS & Total time (4 elec.) & Total time (16 elec.)& Total time (64 elec.) & Total time (256 elec.)\\
    \hline 
     This work & \SI{2262}{\second} & \SI{0.6}{\second} & \SI{2264}{\second} & \SI{2271}{\second} & \SI{2301}{\second} & \SI{2416}{\second} \\
    \hline
     FEM & \SI{81}{\second} & \SI{74}{\second} & \SI{303}{\second} & \SI{1191}{\second} & \SI{4742}{\second} & \SI{18950}{\second} \\
    \hline
    \end{tabular}
\end{center}
\label{table:timecost}
\end{table*}
\section{Simulation Results} \label{sec:numerical}
\subsection{Validation on a canonical model}
Following a well established practice in literature, we benchmark
the accuracy of our new formulation with a standard 3-layer (brain, skull, scalp) spherical model, with normalized radii of \num[round-mode=places,round-precision=2]{0.87}, \num[round-mode=places,round-precision=2]{0.92}, \num{1} and normalized background conductivities of
\SIlist[list-units=single]{1;1/30;1}, respectively.
The skull inhomogeneous conductivity is modeled using the same anisotropic model and as in \cite{dannhauer2011modeling}, with a varying soft bone thickness that is equal to 1 in an arbitrary position $\bm{p}=[0.9, 0, 0]^T$ and decreases linearly to 0 according to the distance to $\bm{p}$.
The white matter anisotropy is modeled in the first sphere by adding \num{30} fibers
with a radius of 0.05, a length of 1.4 and a
longitudinal conductivity that is ten times that of the background. A current dipole acting as the
primary source is placed along the $z$ axis with a dipole moment of $[1, 1, 1]$ in
the $xyz$-coordinate system. The geometry, discretized with an average edge length $h=0.1$, is illustrated in \Cref{fig:sph_svwA}. Leveraging on the reciprocity principle, the hybrid system is directly inverted via LU factorization and backward substitution. A higher resolution FEM solution with quadratic basis functions (\num{3110264} unknowns, $h=0.03$) is used as reference and the accuracy of the hybrid scheme (\num{23358} unknowns) is plotted in \Cref{fig:sph_svw_err} as the relative
$\ell_2$-error of the scalp electric potential, for different dipole excentricities. For comparison, a piecewise-isotropic symmetric BEM solution (\num{12648} unknowns) is computed with
OpenMEEG \cite{gramfort2010openmeeg}, along with an anisotropic FEM
solution (\num{394807} unknowns) obtained with the FieldTrip toolbox \cite{oostenveld2011fieldtrip}. The comparison with the symmetric BEM shows how standard surface modeling, which omits our fiber and volume integral equations, results in important errors (around \SI{20}{\percent}). In contrast, performing slightly better than the completely volumetric FEM solution, our 3-2-1D hybrid formulation exhibits less than \SI{1}{\percent} relative error up to very high source excentricity. This experiment confirms the ability of our hybrid formulation to overcome the model approximations of a standard BEM approach.
\begin{figure}
    \centering
    \trimbox{0pt 0pt 0pt 20pt}{\includegraphics[width=0.8\columnwidth]{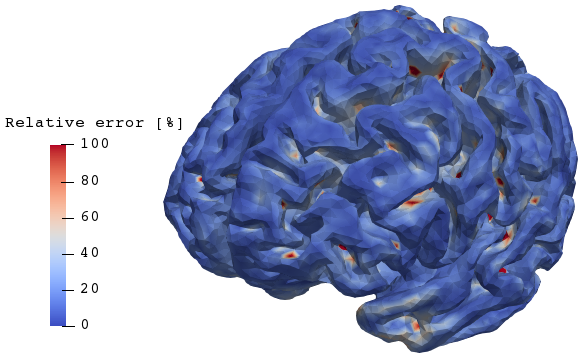}}
    \caption{Cortex map of the relative error between the anisotropic FEM and hybrid solver.}
    \label{fig:relerrFEMvsHybrid}
\end{figure}
\begin{figure*}[ht]
    \centering
    \includegraphics[width=\textwidth]{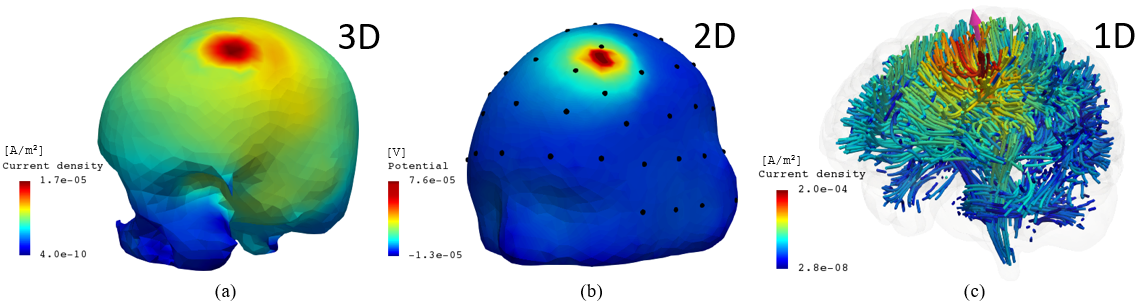}
    \caption{(a) Skull volume currents, (b) scalp surface potential and (c) fiber currents computed with the hybrid formulation. The arrow in (c) represents the cortical dipole and the black dots in (b) the electrode positions.}
    \label{fig:real_mesh}
\end{figure*}
\subsection{Validation of the conductivity model}
The hybrid
solver is also tested and compared with an anisotropic FEM solution on a realistic head model derived from the MRI of a single subject from the Wu-Minn Human Connectome Project database \cite{van2013wu} to highlight its applicability.
Surface and volume meshes were obtained after preprocessing \cite{glasser2013minimal}, segmentation and tessellation \cite{thielscher2015field} of the structural MRI data. The different head compartments were assigned commonly used conductivity values \cite{lozano2013brain}. The skull anisotropy is modeled following the model of \cite{dannhauer2011modeling}, with soft bone thickness defined as an affine function of the local skull thickness. A probabilistic tractography algorithm \cite{tournier2012mrtrix} was applied on the subject's diffusion MRI data to generate a bundle of non-connected streamlines, which was subsequently clustered \cite{garyfallidis2012quickbundles} to obtain a concise fiber map of the white matter. The fiber
radius was adjusted to match a white matter volume of \SI{450}{\centi\meter\cubed}. We applied the procedure described in \Cref{sec:wire_params} with the anisotropic longitudinal and transversal conductivities reported in \cite{NICHOLSON1965386} and obtained the homogenized brain conductivity
\begin{equation}
    \bar{\bar{\sigma}}_e = 
    \begin{bmatrix}
    0.1755 & 0 & 0 \\
    0 & 0.1915 & 0 \\
    0 & 0 & 0.1855
  \end{bmatrix}\,.
\end{equation}
These values are fairly close to the isotropic brain conductivity $\sigma_{brain}=$ \num[round-mode=places,round-precision=2]{0.18} \SI{}{\siemens\per\meter} \cite{lozano2013brain}, and thus confirms that our fiber conductivity model is consistent with the conductivity values reported in literature.
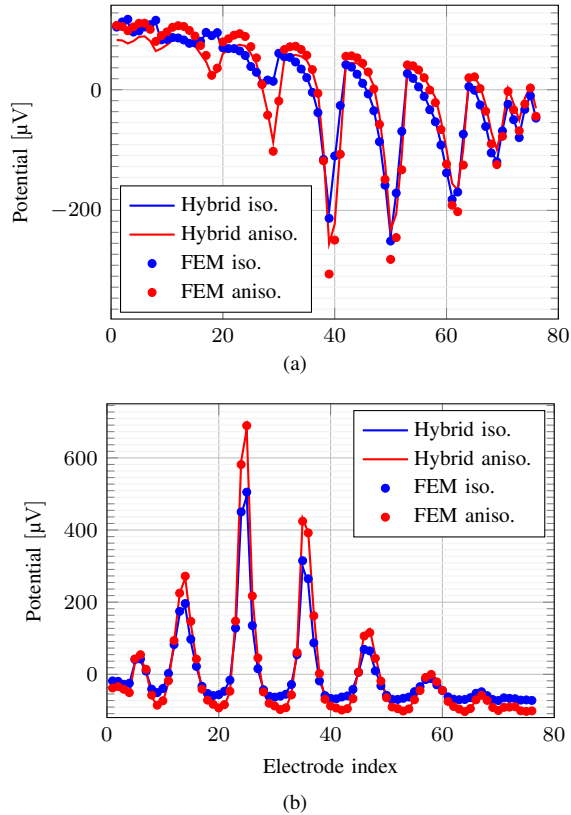
\begin{figure}
    \centering
    \subfloat[]{%
    \trimbox{0pt 10pt 0pt 0pt}{\begin{tikzpicture}
      \begin{axis}[
          width=0.85\columnwidth, 
          height = 0.65\columnwidth,
          grid=both,
          grid style={, draw=gray!10},
          major grid style={,draw=gray!50},
          ylabel=Potential,
          xmin = 0,
          xmax = 80,
          y unit=\si{\micro\volt},
          ymin = -380,
          ymax = 140,
          minor y tick num=10,
          legend style={at={(0.02,0.02)},anchor= south west, font=\footnotesize, legend columns=1},
          legend cell align={left}
        ]
        \addplot [blue, thick]  
        table[x=elec_idx,y=lfs,col sep=comma] {data/pot_6921.txt};
        \addplot [red, thick]
        table[x=elec_idx,y=lfsvw,col sep=comma] {data/pot_6921.txt}; 
        \addplot [blue, mark=*, only marks, mark size=1.5pt]  
        table[x=elec_idx,y=fems,col sep=comma] {data/pot_6921.txt};
        \addplot [red, mark=*, only marks, mark size=1.5pt]
        table[x=elec_idx,y=femsvw,col sep=comma] {data/pot_6921.txt}; 
        \legend{Hybrid iso., Hybrid aniso., FEM iso., FEM aniso.}
      \end{axis}
    \end{tikzpicture}
    }}
    
    \subfloat[]{%
    \begin{tikzpicture}
      \begin{axis}[
          width=0.85\columnwidth,
          height = 0.65\columnwidth,
          grid=both,
          grid style={, draw=gray!10},
          major grid style={,draw=gray!50},
          xlabel=Electrode index,
          ylabel=Potential,
          xmin = 0,
          xmax = 80,
          y unit=\si{\micro\volt},
          ymin = -120,
          ymax = 750,
          minor y tick num=10,
          legend style={at={(0.98,0.98)},anchor= north east, font=\footnotesize},
          legend cell align={left}
        ]
        \addplot [blue, thick]  
        table[x=elec_idx,y=lfs,col sep=comma] {data/pot_11463.txt};
        \addplot [red, thick]
        table[x=elec_idx,y=lfsvw,col sep=comma] {data/pot_11463.txt}; 
        \addplot [blue, mark=*, only marks, mark size=1.5pt]  
        table[x=elec_idx,y=fems,col sep=comma] {data/pot_11463.txt};
        \addplot [red, mark=*, only marks, mark size=1.5pt]
        table[x=elec_idx,y=femsvw,col sep=comma] {data/pot_11463.txt}; 
        \legend{Hybrid iso., Hybrid aniso., FEM iso., FEM aniso.}
      \end{axis}
      \captionsetup{skip=30pt}
    \end{tikzpicture}}
    \caption{Scalp electrode potential resulting from (a) a deep and (b) a shallow source obtained from solving the EEG forward problem with the hybrid integral method (continuous lines) and FEM (dots) and with isotropic (blue) and anisotropic (red) conductivity assumptions.}
    \label{fig:deepshallowpot}
\end{figure}
\subsection{Comparison with an anisotropic full volume FEM model}
An anisotropic FEM solution was obtained from the same dMRI data by computing the diffusion tensor via a least-square fit for each white matter voxel. The conductivity tensor was then derived by following a mean conductivity volume constraint \cite{rullmann2009eeg}. \Cref{fig:relerrFEMvsHybrid} displays the difference between our hybrid and the FEM model, computed as the relative error of the electric potential obtained on a standardized set of \num{76} electrodes (depicted in \Cref{fig:real_mesh}), for each source on a cortical surface. The forward solutions produced by both numerical methods are overall in good agreement, although sources in the cortex sulci exhibit more differences than the shallow gyral sources. This is likely due to the fact that deep sources are more sensitive to the white matter anisotropy, which is modeled differently for both methods. This is further illustrated in \Cref{fig:deepshallowpot} which depicts the scalp electrode potential for a deep and a shallow brain source. The FEM and hybrid forward solution show similar topological and magnitude changes when accounting for the anisotropy. This numerical experiment confirms that the proposed multimodal MRI-based hybrid integral method is consistent with a DTI volume-based anisotropic model of white matter despite their intrinsic modeling differences.
\subsection{Timing comparisons}
All simulations were run on a 24-core 3.00 GHz Intel Xeon E5-2687W v4 CPU with 768 GB of RAM. \Cref{table:timecost} reports the run times of the two methods for different electrode configurations. In total, the FEM system contained \num{1025411} unknowns whereas the hybrid solver had \num{46695}. The proposed method incurs a bigger computational overhead as it must fill an $N^2$ (where $N$ is the number of unknowns) full matrix as opposed to the sparse FEM. Using the reciprocity principle \cite{wolters2004efficient}, for $N_e$ electrodes, the computation of the forward matrix requires the solution of $N_e-1$  RHS. The computational gain becomes apparent for medium to high resolution sensor configurations. Like classical BEM, the $\mathcal{O}(N^2)$ asymptotic complexity in matrix building and storage of the proposed integral equation-based method can furthermore be reduced to linear complexity with fast solvers \cite{bebendorf2000approximation, liu2009fast}.
\section{Conclusion} \label{sec:conclusion}
In this paper we have presented a new solution to the anisotropic EEG forward problem that does not require a full volumetric discretization of the head. The standard boundary integral formulation was coupled with thin volume and wire integral equations that adequately match the non-uniform conductivities of the skull and white matter, respectively. The accuracy of the new BEM-like and anisotropy-adapted formulation was demonstrated on a canonical model while a realistic scenario illustrated its applicability in a clinical environment, in which the patient-specific physiological properties were derived from advanced biomedical imaging techniques.

\bibliographystyle{ieeetr}
\bibliography{biblio}
%
%

\begin{IEEEbiography}[{\includegraphics[width=1in,height=1.25in,clip,keepaspectratio]{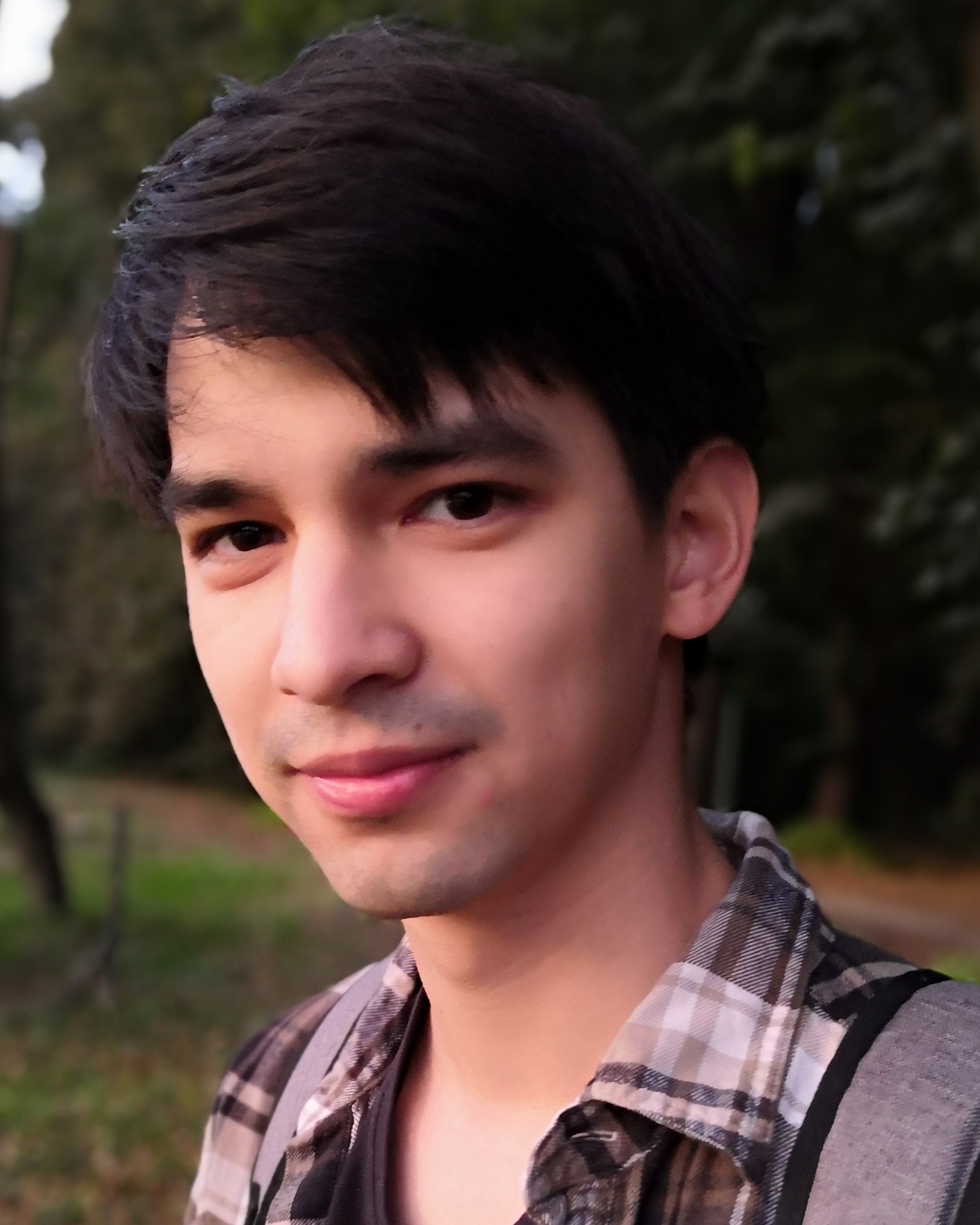}}]{Maxime Y. Monin} (S '18)
received the M.Sc. degree in Information Technologies from IMT Atlantique, Brest, France, and is currently pursuing a Ph.D. with the Department of Electronics and Telecommunications at the Politecnico di Torino, Turin, Italy. His research interests are in computational electromagnetics and their applications for brain imaging.
\end{IEEEbiography}

\begin{IEEEbiography}[{\includegraphics[width=1in,height=1.25in,clip,keepaspectratio]{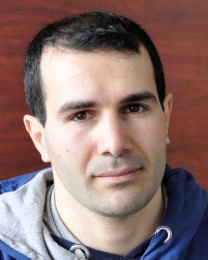}}]{Lyes Rahmouni}
received the B.Sc. in aeronautical engineering and M.Sc. in electronic engineering from the University of Science and Technology Houari Boumediene, Algeria. He has completed a Ph.D. program in computational electromagnetics from IMT Atlantique, France in 2016, followed by two years of postdoctoral position at the same grande \'ecole. He is currently holder of a Research fellowship at Politecnico di Torino, Italy. His research interests focus on integral equation methods and their fast solutions.
\end{IEEEbiography}

\begin{IEEEbiography}[{\includegraphics[width=1in,height=1.25in,clip,keepaspectratio]{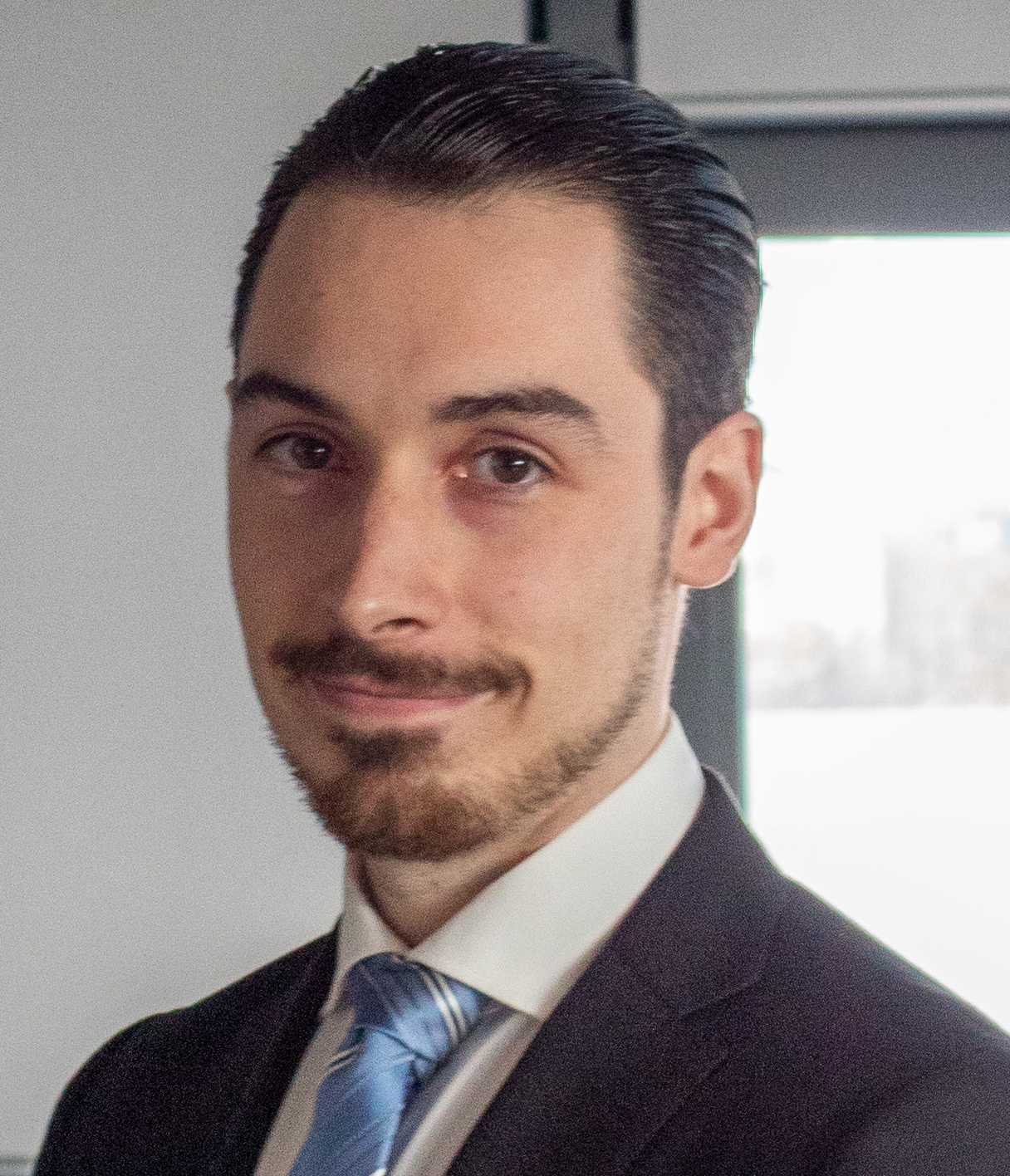}}]{Adrien Merlini}
(S '16, M '19) received the M.Sc. Eng. degree from the
\'Ecole Nationale Sup\'erieure des T\'el\'ecommunications de Bretagne (T\'el\'ecom
Bretagne), France, in 2015 and received the Ph.D. degree from the \'Ecole
Nationale Sup\'erieure Mines-T\'el\'ecom Atlantique (IMT Atlantique), France,
in 2019. Between 2018 and 2019, he was a visiting Ph.D. student at the
Politecnico di Torino, Italy, which he then joined as a Research
Associate. Since the end of 2019, he is an Associate Professor with the
Microwave Department of IMT Atlantique, Brest, France.

His research interests include preconditioning and acceleration of
integral equation solvers for electromagnetic simulations and their
application in brain imaging.
\end{IEEEbiography}

\begin{IEEEbiography}[{\includegraphics[width=1in,height=1.25in,clip,keepaspectratio]{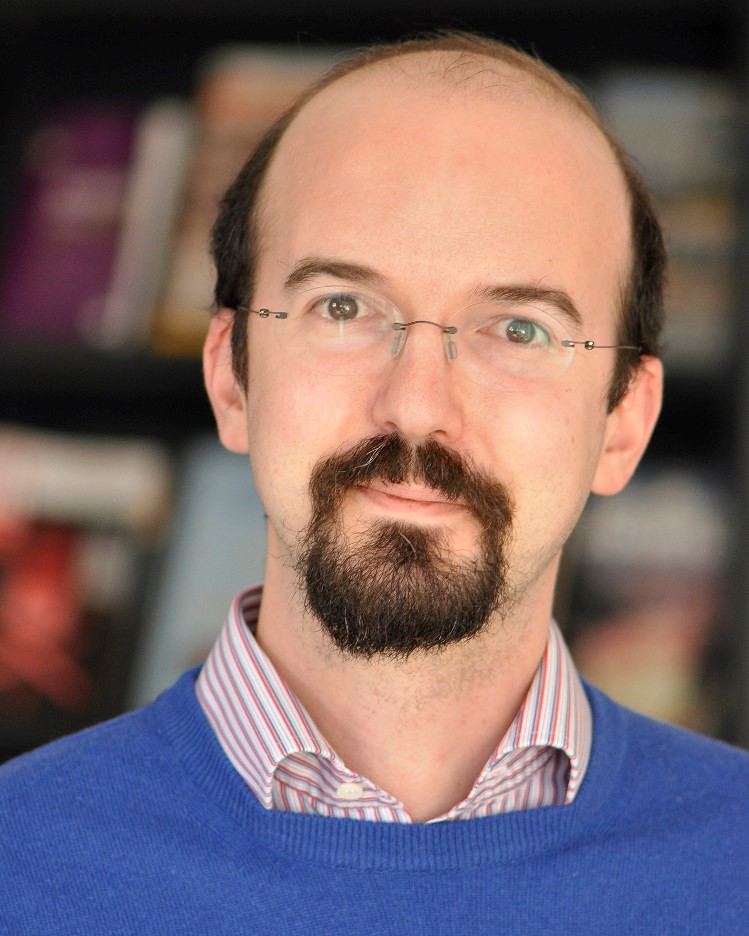}}]{Francesco P. Andriulli}
(S '05, M '09, SM '11)  received the Laurea in electrical engineering from the Politecnico di Torino, Italy, in 2004, the MSc in electrical engineering and computer science from the University of Illinois at Chicago in 2004, and the PhD in electrical engineering from the University of Michigan at Ann Arbor in 2008. From 2008 to 2010 he was a Research Associate with the Politecnico di Torino. From 2010 to 2017 he was an Associate Professor (2010-2014) and then Full Professor with the \'Ecole Nationale Sup\'erieure Mines-T\'el\'ecom Atlantique (IMT Atlantique, previously ENST Bretagne), Brest, France. Since 2017 he has been a Full Professor with the Politecnico di Torino, Turin, Italy. His research interests are in computational electromagnetics with focus on frequency- and time-domain integral equation solvers, well-conditioned formulations, fast solvers, low-frequency electromagnetic analyses, and modeling techniques for antennas, wireless components, microwave circuits, and biomedical applications with a special focus on Brain Imaging.

Prof. Andriulli was the recipient of the best student paper award at the 2007 URSI North American Radio Science Meeting.  He received the first place prize of the student paper context of the 2008 IEEE Antennas and Propagation Society International Symposium. He was the recipient of the 2009 RMTG Award for junior researchers and was awarded two URSI Young Scientist Awards at the International Symposium on Electromagnetic Theory in 2010 and 2013 where he was also awarded the second prize in the best paper contest. He also received the 2015 ICEAA IEEE-APWC Best Paper Award. In addition, he co-authored with his students and collaborators other three first prize conference papers (EMTS 2016, URSI-DE Meeting 2014, ICEAA 2009), a second prize conference paper (URSI GASS 2014), a third prize conference paper (IEEE-APS 2018), two honorable mention conference papers (ICEAA 2011, URSI/IEEE-APS 2013) and other three finalist conference papers (URSI/IEEE-APS 2012, URSI/IEEE-APS 2007, URSI/IEEE-APS 2006). Moreover, he received the 2014 IEEE AP-S Donald G. Dudley Jr. Undergraduate Teaching Award, the triennium 2014-2016 URSI Issac Koga Gold Medal, and the 2015 L. B. Felsen Award for Excellence in Electrodynamics. 

Prof. Andriulli is a member of Eta Kappa Nu, Tau Beta Pi, Phi Kappa Phi, and of the International Union of Radio Science (URSI). He is the Editor in Chief of the IEEE Antennas and Propagation Magazine, he serves as a Track Editor for the IEEE Transactions on Antennas and Propagation, and as an Associate Editor for the IEEE Antennas and Wireless Propagation Letters, IEEE Access, URSI Radio Science Letters and IET-MAP. He is the PI of the ERC Consolidator Grant ``321''.
\end{IEEEbiography}

\vfill






\end{document}